\documentclass{jfm}
\usepackage[authoryear]{natbib}
\usepackage{graphicx}
\usepackage{epstopdf, epsfig}
\usepackage[usenames,dvipsnames,svgnames,table]{xcolor}
\usepackage[lofdepth,lotdepth]{subfig}

\makeatletter
\def\mathcolor#1#{\@mathcolor{#1}}
\def\@mathcolor#1#2#3{
  \protect\leavevmode
  \begingroup
    \color#1{#2}#3
  \endgroup
}
\makeatother

\shorttitle{Boundary streaming by internal waves}
\shortauthor{A. Renaud and A. Venaille}

\title{Boundary streaming by internal waves}

\author{A. Renaud\corresp{\email{antoine.renaud@ens-lyon.fr}},  A. Venaille}

\affiliation{Univ Lyon, Ens de Lyon, Univ Claude Bernard, CNRS, Laboratoire de Physique, F-69342 Lyon, France}

\begin{document}

\maketitle

\begin{abstract}
Damped internal wave beams in stratified fluids have long been known to generate strong mean flows through a mechanism analogous to acoustic streaming. While the role of viscous boundary layers in acoustic streaming has thoroughly been addressed, it remains largely unexplored in the case of internal waves. Here we compute the mean flow generated close to an undulating wall that emits internal waves in a viscous, linearly stratified two\--dimensional Boussinesq fluid. Using a quasi\--linear approach, we demonstrate that the form of the boundary conditions dramatically impacts the generated boundary streaming. In the no-slip scenario, the early time Reynolds stress divergence within the viscous boundary layer is much stronger than within the bulk while also driving flow in the opposite direction. Whatever the boundary condition, boundary streaming is however dominated by bulk streaming at large time. Using a WKB approach, we investigate the consequences of adding boundary streaming effects to an idealised model of wave-mean flow interactions known to reproduce the salient features of the quasi\--biennial oscillation. The presence of wave boundary layers has a quantitative impact on the flow reversals.
\end{abstract}


\section{Introduction}

Internal gravity waves play a crucial role in the dynamics of atmospheres and oceans by redistributing energy and momentum \citep{Sutherland2010}. In particular, strong mean flows can be generated by non\--linear effects within internal wave beams \citep{Lighthill1978}, a phenomenon analogous to acoustic streaming {\citep{Riley2001,eckart1948vortices}.  Internal wave streaming is central to the quasi\--biennial oscillation of equatorial zonal winds in the equatorial stratosphere \citep{Baldwin2001}. The salient features of this robust phenomenon have been reproduced in a celebrated laboratory experiment \citep{Plumb1978} and in direct numerical simulations \citep{wedi2006direct}.  Since then, other instances of internal wave streaming have been reported in various experimental and numerical configurations:  \cite{Semin2016} used a quasi two\--dimensional experimental setting similar to \citet{Plumb1978} to describe internal wave streaming in the absence of flow reversal; \citet{Grisouard2012,Venaille2012,Akylas2015} showed that three\--dimensional effects lead to vortical streaming in the domain bulk.}
However, those previous studies have not addressed the role of viscous boundary layers and their potential implications for the generation of mean flows confined to the boundary. This contrasts with acoustic waves which have long been known to produce strong mean flows within their viscous boundary layers \citep{rayleigh1884circulation,Nyborg1958}. Boundaries are essential to the generation of the waves in laboratory experiments  \citep{Gostiaux2006} or numerical models \citep{Legg2014}, and to energy focusing \citep{Maas1997}. In the atmosphere and oceans, internal gravity waves are often generated through the interaction between a mean flow and a solid boundary (orography in the atmosphere, bathymetry in the oceans). Viscous effects are negligible at those geophysical scales, but  numerical simulations of these flows are usually performed with larger effective turbulent viscosities. It is therefore crucial to understand the effect of viscous boundary layers.  Viscous internal wave beams generated by boundaries have been extensively studied \citep{Voisin2003}, together with their consequences on the bulk energy budget of numerical ocean models \citep{Shakespeare2017}. The role of viscous boundary layers has been addressed by \citet{Maas2016} to close the energy budget of internal wave attractors;  \citet{Chini2003}  described the viscous boundary layers in the case of Klemp and Durran boundary conditions; \citet{passaggia2014response} studied the structure of a stratified boundary layer over a tilted bottom with a small stream\--wise undulation.  {The effect of the viscous boundary layers on the mean flow is not discussed in those works. By contrast, \citet{Grisouard2015,Grisouard2016} carried out full nonlinear simulations of internal wave reflections and showed the existence of strong mean flows induced by the waves in the vicinity of a reflecting boundary. They also showed the importance of the wave boundary layers in the energy budget of the mean flow. This provides a strong incentive to revisit the mean flow generation associated with internal gravity wave boundary layers.}

Here, using a two\--dimensional and quasi\--linear framework, we compute the mean flow generated by internal gravity waves close to a boundary, paying particular attention to the role of boundary conditions. The importance of changing the boundary condition in numerical models of internal wave dynamics close to bottom topography has been noticed in previous work related to mixing and wave dissipation \citep{Nikurashin2010}. We will show that changing boundary conditions also substantially affects wave\--driven mean flows.  The quasi\--linear approach is introduced in section \ref{sec:ZSWM}. The structure of the viscous linear waves, their induced Reynolds stress divergences and the consequences for mean flow generation are discussed in section \ref{sec:ViscIW}. An application to an idealised model of a quasi\--biennial oscillation analogue is presented in section \ref{sec:MeanFlow}. A WKB treatment of the problem is provided in appendix \ref{sec:WKB}. 

\section{Internal gravity wave-mean flow interactions with zonal symmetry}\label{sec:ZSWM}

We consider a fluid within a two\--dimensional domain, periodic in the zonal $x$\--direction with period $L$ and semi\--infinite in the vertical $z$\--direction. The bottom boundary is a vertically undulating line located on average at $z=0$. The fluid is considered incompressible, Boussinesq, viscous  with viscosity $\nu$ and linearly stratified with buoyancy frequency $N$. For the sake of simplicity, we ignore any buoyancy diffusion process. This approximation is relevant for experimental configurations where the stratification agent is salt,  given the low diffusivity $\kappa=\nu/1000$,  {but it does not apply to the atmosphere and the ocean, where turbulent viscosity and diffusivity have the same order of magnitude.} 

{Throughout this work, we solely consider monochromatic waves. Let us introduce the typical zonal wave number $k=2\pi/L$, angular frequency $\omega$ and amplitude of the bottom undulation $h_{b}$. There are three independent dimensionless numbers in the problem. The Froude number $F\!r=\omega/N$ controls the angle of propagation of the wave. The wave Reynolds number  $Re=\omega/\left(k^{2}\nu\right)$ controls the viscous damping and the viscous boundary layer thickness of the wave field. When considering the lee wave generation case, this wave Reynolds number scales as $UL/\nu$, where $U$ is the typical mean zonal velocity. The third parameter is the dimensionless amplitude of the wave $\epsilon=h_{b}k$. It corresponds to the typical slope of the bottom boundary, controlling the linearity of the wave. In numerical simulations, an additional aspect ratio $r=kH$ and a wave P\'eclet number $\omega / (k^2 \kappa)$ have to be taken into account, because the domain has a finite height $H$, and because it includes a buoyancy diffusivity $\kappa$. Both parameters will be much larger than one in the numerical simulations presented in this paper, and we will assume that they do not play a significant role in this limit. We use $k^{-1}$, $\omega^{-1}$ as reference length and time for the space\--time coordinates, $c=\omega/k$ as a reference velocity, $N^{2}/k$ as a reference buoyancy, and write the dynamical equation in a dimensionless form}

\begin{equation}\label{eq:BoussModel}
	\left\{\begin{array}{l l}
		\p_{t}\mathbf{u}+\left(\mathbf{u}\cdot\nabla\right)\mathbf{u}&=-\nabla p +{F\!r^{-2}b\mathbf{e}_{z}+Re^{-1}\nabla^{2}\mathbf{u}}\\
		\p_{t}b+\mathbf{u}\cdot\nabla b+{w} &=0\\
		\nabla\cdot\mathbf{u} &=0
	\end{array}\right.   {,}
\end{equation}
where $\mathbf{u}=\left(u,w\right)$ is the two\--dimensional velocity, $p$ the renormalised pressure, $b$ the buoyancy anomaly, {$\mathbf{e}_z$ the unit vector of the vertical direction pointing upward}, and $\nabla^{2}=\p_{xx}+\p_{zz}$ the standard Laplacian operator.

Previous studies in the context of acoustic streaming have investigated the effect of changing boundary conditions on mean flow properties \citep{Xie2014}. In this paper devoted to internal wave streaming,  we discuss two different bottom boundary conditions on $z={\epsilon} h\left(x,t\right)$:
\begin{equation}\label{eq:BC}
	\mbox{free\--slip:}\quad w={\epsilon\left(\p_{t}h+u\p_{x}h\right)}\;,\;\mathsfbi{G}\left[\mathbf{n}_{h}\right]\cdot\mathbf{n}^{\bot}_{h}=0\quad;\quad\mbox{no\--slip:}\quad \mathbf{u}={\epsilon}\p_{t}h\mathbf{e}_{z}   {,} 
\end{equation}
where $\mathbf{n}_{h}=\nabla\left(z-{\epsilon} h\left(x,t\right)\right)$ is a local normal vector of the bottom boundary, $\mathbf{n}_{h}^{\bot}$ a local tangent vector and $\mathsfbi{G}$ the {velocity gradient tensor} ($G_{ij}=\p_{j}u_{i}$).  {This free-slip condition is the one implemented in the numerical model considered in this paper \citeyearpar[see MITgcm user's manual][]{MitDOC}. It is equivalent to the stress-free condition when boundary curvature can be neglected.  In the stress-free case, $\mathsfbi{G}$ is replaced by its symmetric part only. Regarding the boundary streaming, we checked the discrepancies between stress-free and our free-slip condition arise only in non-hydrostatic regimes of internal waves. Therefore, in most practical cases, the results obtained by considering the free-slip condition  (\ref{eq:BC}) will also be relevant for numerical simulation using the stress-free condition. Furthermore, we require all gradients with respect to $z$ to vanish as $z\to\infty$.}

When considering a progressive pattern ($h\left(x,t\right)=h\left(x-t\right)$) in (\ref{eq:BC}), a Galilean change of reference  {yields} the case of lee-wave generation by a depth-independent mean flow passing over bottom topography. Then, the free\--slip bottom boundary condition for the generation of lee waves  {obviates the need to treat the near-bottom} critical layer induced by a more realistic no\--slip condition \citep{passaggia2014response}. Regarding the free\--slip condition, the predictions will be compared against direct numerical simulations of monochromatic lee waves generation using the MIT global circulation model \citep{adcroft1997} which specifically uses our definition for the free\--slip condition. The no\--slip boundary condition in (\ref{eq:BC}) is relevant to model the generation of internal gravity waves in laboratory experiments using vertically oscillating bottom membranes \citep{Plumb1978,Semin2016} or a system of plates and camshafts \citep{Gostiaux2006}. We will, however, consider limiting cases where the viscous boundary layer is larger than the boundary height variations, which is not always the case in actual experiments.\\

We decompose any field $\phi$ into a mean flow  part $\overline{\phi}$ and a wave part $\phi^{\prime}$ using the zonal averaging procedure  \citep[see][]{Buhler2009}:
\begin{equation}\label{eq:WMDef}
	\overline{\phi}\left(z,t\right)=\frac{1}{{2\pi}}\int_{0}^{{2\pi}}\mathrm{d}x
	\;\phi\left(x,z,t\right),\quad\phi^{\prime}=\phi-\overline{\phi}.
\end{equation}
The averaging of the zonal momentum equation in (\ref{eq:BoussModel}) leads to the mean flow evolution equation: 
\begin{equation}\label{eq:Evol_MeanU}
	\p_{t}\overline{u}=-\p_{z}\overline{u^{\prime}w^{\prime}}+{Re^{-1}}\p_{zz}\overline{u}.
\end{equation}
{The source of streaming is the divergence of the Reynolds stress $-\p_{z}\overline{u^{\prime}w^{\prime}}$. To compute this term, we subtract} the averaged equations from (\ref{eq:BoussModel}) and {linearise the result assuming $\left(u^{\prime},w^{\prime},b^{\prime},p^{\prime}\right)=O\left(\epsilon\right)$ with $\epsilon\ll 1$.} 

{At this stage, we assume that $\left|\overline{u}\right|\ll 1$. Starting from a state of rest, at early times of its evolution, the mean flow is weak, which justifies this assumption.  {At later times, the feedback of the mean flow on the wave can no longer be ignored  {\citep{Akylas2015,fan2018interaction}}, as will be discussed in more detail in section \ref{sec:MeanFlow} (see also equation (\ref{eq:LinWithMean}) in appendix \ref{sec:WKB})}.  This  case without feedback from the mean flow leads to homogeneous wave equations, which provides a simple framework to describe essential features of boundary streaming:} 
\begin{equation}\label{eq:GlobalFluct}{
	\left\{
	\begin{array}{ll}
		\p_{t}u^{\prime}+\p_{x}p^{\prime}-Re^{-1}\nabla^{2} u^{\prime}&=0\  \\
		\p_{t}w^{\prime}+\p_{z}p^{\prime}-F\!r^{-2}b^{\prime}-Re^{-1}\nabla^{2} w^{\prime}&=0\ \\
		\p_{t}b^{\prime}+w^{\prime}& =0\ \\
		\p_{x}u^{\prime}+\p_{z}w^{\prime} & =0 \ 
	\end{array}
	\right.}   {.}
\end{equation}
The coupled equations (\ref{eq:Evol_MeanU}) and (\ref{eq:GlobalFluct}) form a quasi\--linear model for the interaction between boundary generated viscous waves and the zonal mean flow. The Reynolds stress {divergence}, $-\p_{z}\overline{u'w'}$, at the origin of streaming, is the only non\--linear term remaining in the problem. It acts as a forcing term and is computed from the wave field.

We perform the wave\--mean decomposition on the boundary conditions (\ref{eq:BC}) {and we linearise the result assuming as above a wave amplitude of order $\epsilon$} on an asymptotically flat boundary at $z=0$:
\begin{equation}\label{eq:BC_Lin}
	\begin{array}{llll}
		\mbox{free\--slip:}&\quad\left\{
		\begin{array}{ll}
			\p_{z}\overline{u}&=0\\
			w^{\prime}-\partial_{t}h&=0\\
			\p_{z}u^{\prime}&=0
		\end{array}\right.&\quad;\quad
		\mbox{no\--slip:}&\quad\left\{
		\begin{array}{ll}
			\overline{u}&=0\\
			w^{\prime}-\partial_{t}h&=0\\
			u^{\prime}&=0
		\end{array}\right.
	\end{array}.
\end{equation}
In the free\--slip case, the Reynolds stress {divergence} vanishes at the bottom ($\p_{z}\overline{u^{\prime}w^{\prime}}|_{z=0}=0$) while, in the no\--slip case,   the Reynolds stress  itself vanishes at the bottom ($\overline{u^{\prime}w^{\prime}}|_{z=0}=0$). Given that $\overline{u^{\prime}w^{\prime}}|_{z{\to}\infty}=0$ for damped waves, the integrated streaming {in the no-slip case} has to be zero: $\int_{0}^{\infty}\p_{z}\overline{u^{\prime}w^{\prime}}\,\mathrm{d}z=0$. Consequently, all the streaming far from the bottom boundary has to be compensated for by an opposite boundary streaming.

\section{From viscous waves to boundary streaming}\label{sec:ViscIW}

\subsection{Viscous internal gravity waves}

We describe in this section the detailed structure of the Reynolds stress {divergences} for both the free-slip and the no-slip boundary conditions, when the mean flow can be neglected. Inserting the ansatz $\left(u^{\prime},w^{\prime},b^{\prime},p^{\prime}\right)=\Re\left[\left(\tilde{u},\tilde{w},\tilde{b},\tilde{p}\right)e^{i\left({x+mz-t}\right)}\right]$ into equation (\ref{eq:GlobalFluct}) leads to the dispersion relation for viscous internal gravity waves, expressed here as 
\begin{equation}
	m^{2}=\frac{iRe}{2}\left(1\pm\sqrt{1+\frac{4i}{F\!r^{2}Re}}\right)-1. \label{eq:DispersionRel}
\end{equation}
Among the four possible solutions for $m$, we retain only the two upward propagating ones, by discarding the solutions with a negative imaginary part. 
To simplify the discussion,  { it will be useful to express }  these solutions in the asymptotic regime $F\!r^{2}Re\gg1$, followed by $F\!r\ll 1$:
\begin{equation}\label{eq:m_Val}{
	\left\{
	\begin{array}{ll}
 		m_{w}  &= -1/F\!r +i/\left(2L_{Re}\right)+o\left(\left(Re F\!r^{3}\right)^{-1}\right)\\
		m_{bl} &= \left(1+i\right)/\delta_{Re}+o\left(Re^{1/2}\right)
	\end{array}\right.   {,}}
\end{equation}
with
\begin{equation}\label{eq:LengthScales}{
		L_{Re}   =Re F\!r^{3}\quad\mathrm{and}\quad\delta_{Re}  =\sqrt{2/Re} \ .}
\end{equation}

The solution $m_{w}$ corresponds to the propagating solution converging toward the inviscid solution in the limit ${Re\to \infty}$. $L_{Re}$ is the damping length\--scale of the wave-beam, scaling {linearly with the wave Reynolds number}. The solution $m_{bl}$ corresponds to the wave boundary layer. The boundary layer thickness, given by $\delta_{Re}$, scales {as $Re^{-1/2}$} as in the classical case of a horizontally oscillating flat boundary. This last solution is needed to match the propagating solution with the viscous boundary conditions and is analogous to the one discussed in acoustic boundary streaming \citep{Nyborg1958}. Importantly the ratio $L_{Re}/\delta_{Re}$ diverges in the limit $F\!r^2 Re \rightarrow +\infty$.  {This limit, therefore, allows for a clear separation between bulk and boundary effects.} 

The viscous internal-wave dispersion relation has already been extensively studied. \citet{Chini2003} considered a finite Prandtl number, which provides an additional branch of boundary layer solutions associated with the diffusion operator in the buoyancy equation. They also gave asymptotic expansions for large Reynolds number. \citet{Grisouard2016} considered the effect of a Coriolis force, which provides an additional branch of boundary layers.  {Although rotation, buoyancy diffusion, and their associated boundary layer solutions undoubtedly impact boundary streaming, we do not consider these additional effects, to simplify the presentation.} 

In the case of a progressive sine\--shaped bottom undulations, $h\left(x,t\right)={\Re\left[e^{i\left(x-t\right)}\right]}$, the general expression of the wave field is given by the linear combination of a propagating  {($w$)} and a boundary layer  {($bl$)} part
\begin{equation}\label{eq:WaveGeneral}
	\left[u^{\prime},w^{\prime},b^{\prime},p^{\prime}\right] =  \Re\left\{{\left(\phi_{w}\mathbf{P}\!\left[m_{w}\right]
	e^{i m_{w}z}+\phi_{bl}\mathbf{P}\left[m_{bl}\right]e^{i m_{bl}z}\right)e^{i\left(x-t\right)}}\right\},
\end{equation}
with 
\begin{equation}
	\mathbf{P}\!\left[m\right]=\left[{1,-m^{-1},iF\!r^{-2}m^{-1},F\!r^{-2}\left(1+m^{2}\right)^{-1}}\right].\label{eq:PolarWave}
\end{equation}
$\mathbf{P}\!\left[m\right]$ is the polarisation of the wave obtained from (\ref{eq:GlobalFluct}), $\left(m_{w},m_{bl}\right)$ are given in equation (\ref{eq:m_Val}), and $\left(\phi_{w},\phi_{bl}\right)$ are scalars determined by the boundary conditions (\ref{eq:BC_Lin}):
\begin{equation}\label{eq:BC_Fi}
	\begin{array}{llll}
		\mbox{free\--slip:}&\left\{
		\begin{array}{ll}
			\phi_{w}&=i{\epsilon}\frac{m_{w}m_{bl}^{2}}{m_{bl}^{2}-m_{w}^{2}}\\
			\phi_{bl}&=i{\epsilon}\frac{m_{bl}m_{w}^{2}}{m_{w}^{2}-m_{bl}^{2}}
		\end{array}\right.;
		& 
		\mbox{no\--slip:}&\left\{
		\begin{array}{ll}
			\phi_{w}&=i{\epsilon}\frac{m_{w}m_{bl}}{m_{bl}-m_{w}}\\
			\phi_{bl}&=i{\epsilon}\frac{m_{w}m_{bl}}{m_{w}-m_{bl}}
		\end{array}\right.
	\end{array}.
\end{equation}
  
\begin{figure}
	\centering \includegraphics[width=1\columnwidth]{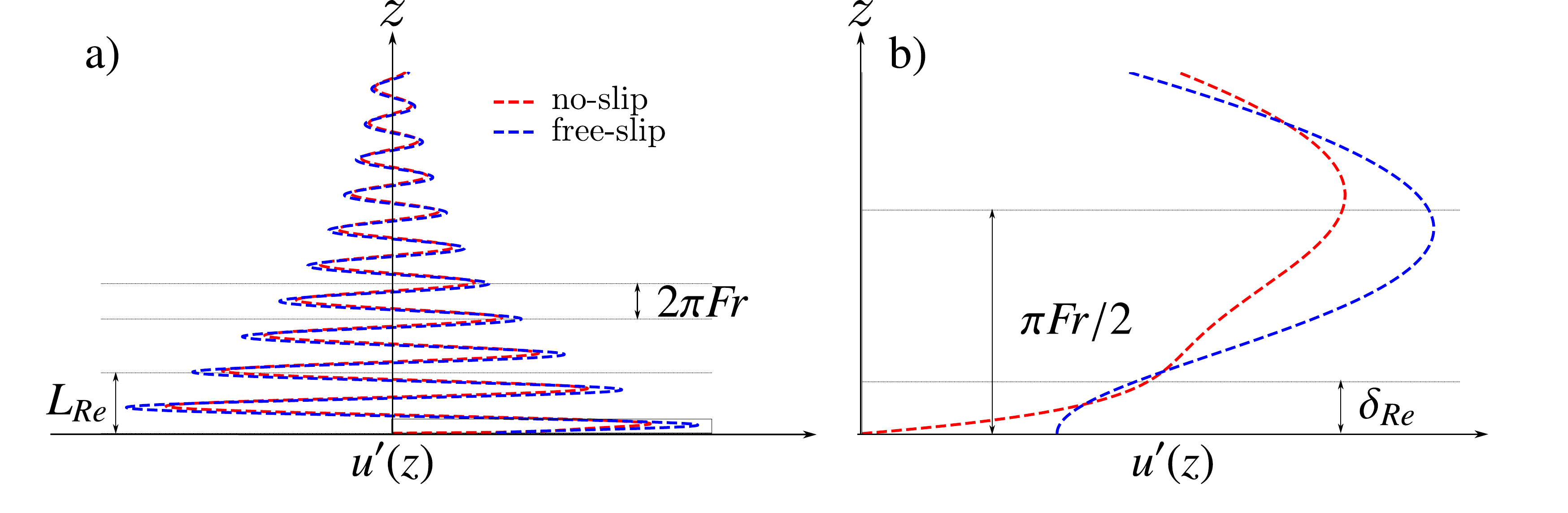}
	\caption{\label{fig:IV_Illus}a) Example of a linear computation of the vertical profile of the fully established wave field, $u^{\prime}$, in the absence of a mean flow, with the free\--slip ({blue}) and  no\--slip (red) boundary conditions. b) Zoom on the boundary layer of the wave. The wave damping length, $L_{Re}$, and the boundary layer thickness, $\delta_{Re}$, are represented on the graph along with the inviscid vertical wavelength, $\lambda_{z}=2\pi F\!r$.}
\end{figure}
The generic vertical profiles of the wave field $u^{\prime}$ are drawn in figure \ref{fig:IV_Illus} for both boundary conditions. Most of the differences between the two profiles are located in the boundary layer close to the bottom. We will see that these different profiles  lead to very different boundary streaming behaviours, by computing the Reynolds stress {divergence} of the corresponding wave fields. 

\subsection{Reynolds stress divergence}

The Reynolds stress $\overline{u^{\prime}w^{\prime}}$ is composed of cross terms involving both the propagative and the boundary layer contributions. In the limit of small viscosity, the ``self\--interaction'' of the propagating contribution decreases exponentially over a scale $L_{Re}$. This corresponds to bulk streaming. All the other terms involve a pairing with the boundary layer contribution that decay exponentially over the scale $\delta_{Re}$. The sum of these terms induces the boundary streaming. We thus decompose the Reynolds stress into a bulk  and a boundary term  {
\begin{equation}\label{eq:BulkBL_decomp}
	\overline{u^{\prime}w^{\prime}}\left(z\right)=F_{w}
	\left(z\right)+F_{bl}\left(z\right).
\end{equation}}

{In the remainder of this section, the quasi\--linear computations will be performed by using the exact solutions of (\ref{eq:DispersionRel}). In order to get insights on the basic differences between the free-slip and the no-slip case, it is useful, however, to estimate the Reynolds stress by using the asymptotic expression (\ref{eq:m_Val}) for both boundary conditions in (\ref{eq:BC_Lin}):  {
\begin{equation}\label{eq:BC_RS}{
	\begin{array}{ll}
		\mbox{free\--slip:}&\quad\left\{
		\begin{array}{ll}
			F_{w}\left(z\right)    &= \frac{\epsilon^{2}}{2F\!r}\exp\left\{-\frac{z}{L_{Re}}\right\}\\
			F_{bl}\left(z\right)   &= \frac{\epsilon^{2}}{F\!r^2 2\sqrt{2Re}}\exp\left\{-\frac{z}{\delta_{Re}}\right\}\left(\sin\frac{z}{\delta_{Re}}+\cos\frac{z}{\delta_{Re}}\right)
		\end{array}\right.\\
		& \\
		\mbox{no\--slip:}&\quad\left\{
		\begin{array}{ll}
			F_{w}\left(z\right)    &= \frac{\epsilon^{2}}{2F\!r}\exp\left\{-\frac{z}{L_{Re}}\right\}\\
			F_{bl}\left(z\right)   &= -\frac{\epsilon^{2}}{2F\!r}\exp\left\{-\frac{z}{\delta_{Re}}\right\}\cos\frac{z}{\delta_{Re}}
		\end{array}\right.
	\end{array}.}
\end{equation}}

The bulk  {Reynolds stress $F_{w}$} has the same expression at leading order for both the free\--slip and the no\--slip case. The difference {lies} in the boundary {'s Reynolds stress expression $F_{bl}$}.  {Similarly, the asymptotic expressions  of the streaming body forces are:}

\begin{equation}\label{eq:BC_RSB}{
	\begin{array}{ll}
		\mbox{free\--slip:}&\quad\left\{
		\begin{array}{ll}
			-\p_{z}F_{w}\left(z\right)    &= \frac{\epsilon^{2}}{2F\!r^{4}Re}\exp\left\{-\frac{z}{L_{Re}}\right\}\\
			-\p_{z}F_{bl}\left(z\right)   &= \frac{\epsilon^{2}}{2F\!r^2}\exp\left\{-\frac{z}{\delta_{Re}}\right\}\sin\frac{z}{\delta_{Re}}
		\end{array}\right.\\
		& \\
		\mbox{no\--slip:}&\quad\left\{
		\begin{array}{ll}
			-\p_{z}F_{w}\left(z\right)    &= \frac{\epsilon^{2}}{2F\!r^{4}Re}\exp\left\{-\frac{z}{L_{Re}}\right\}\\
			-\p_{z}F_{bl}\left(z\right)   &= -\frac{\epsilon^{2}\sqrt{Re}}{2F\!r \sqrt{2}}\exp\left\{-\frac{z}{\delta_{Re}}\right\}\left(\cos\frac{z}{\delta_{Re}}+\sin\frac{z}{\delta_{Re}}\right)
		\end{array}\right.
	\end{array}.}
\end{equation}
}

\begin{figure}
	\centering\includegraphics[width=0.75\columnwidth]{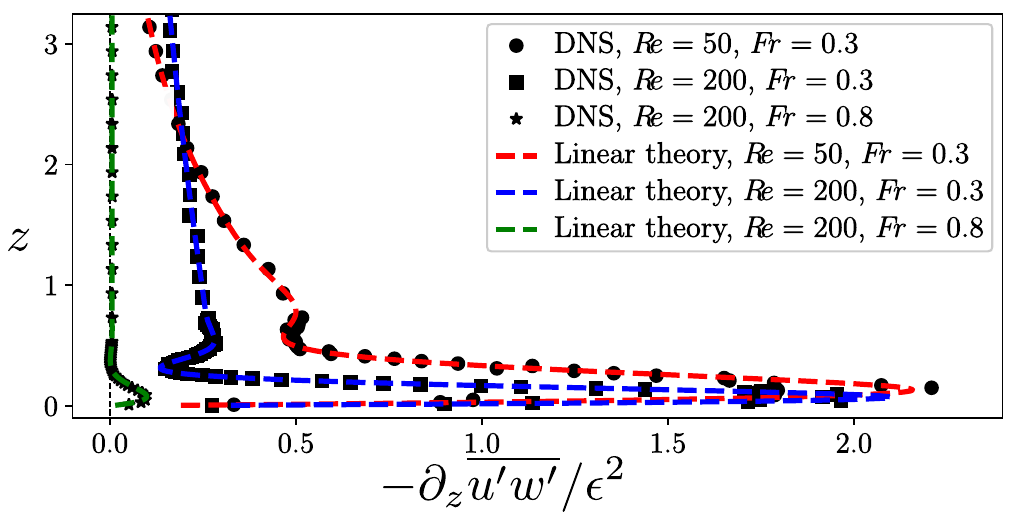}
	\caption{\label{fig:MIT}Plot of the vertical profile of the Reynolds stress {divergence} in the absence of mean flow ($\overline{u}=0$) considering the free\--slip boundary condition for different couples $\left(Re,F\!r\right)$. The markers plots come from high\--resolution direct numerical simulations (DNS) while the dashed lines plots come from the full linear theory without mean flow. The other dimensionless parameters for the simulation are $\epsilon=0.01$ (wave amplitude) and $r=6L_{{Re}}$ (domain aspect ratio); the resolution is $\Delta x=\Delta z=\delta_{{Re}}/50$ ; the grid is stretched  above $z=6L_{Re}$ to avoid wave reflection; the simulated data have been smoothed over ten time steps of the simulation {to get rid of the fast motion coming from surface waves present in the numerical model}.}
\end{figure}

 {In the free\--slip case, the boundary forcing amplitude  does not depend on the wave Reynolds number at leading order, only its e-folding height does.} This amplitude decreases with the Froude number. This effect can be seen {in} figure \ref{fig:MIT} where the free\--slip Reynolds stress {divergence} $\p_{z}\overline{u^{\prime}w^{\prime}}$ is plotted for three different values of Reynolds and Froude numbers. These quasi\--linear calculations are successfully compared to high resolution direct numerical simulations of the established wave pattern generated by a depth-independent flow above a sine\--shaped topography in a linearly stratified fluid.

\begin{figure}
	\centering\includegraphics[width=1.\linewidth]{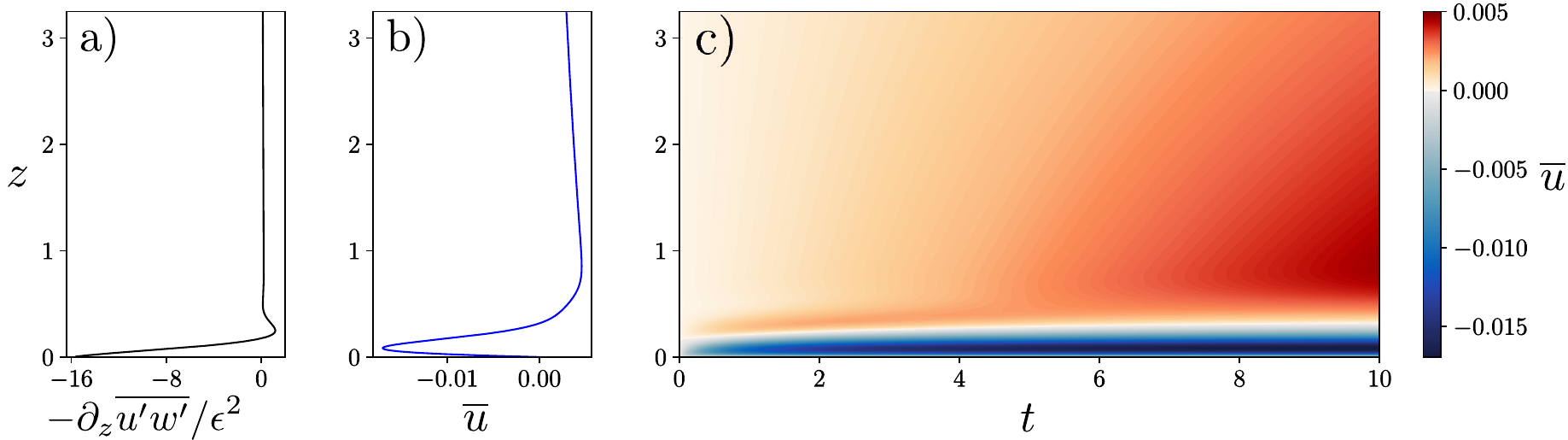}
	\caption{\label{fig:LAB} a) Plot of the vertical profile of the Reynolds stress {divergence} for the no\--slip boundary condition computed using the full linear theory without mean flow. b) Plot of the vertical profile of the mean flow at $t=10$ computed using the quasi\--linear model for the no\---slip boundary condition. c) Hovm\"oller diagrams of the mean flow {, $\overline{u}\left(z,t\right)$, computed using the quasi\--linear theory for the scenario in which the lower boundary condition is no\--slip.} The parameters are $Re=200$, $F\!r=0.3$ and {$\epsilon=0.005$}. }
\end{figure}

 {In the no\--slip case, boundary forcing is opposite (and much stronger) than the bulk forcing, as shown in figure \ref{fig:LAB}-a. The underlying reason is the vanishing of the integral of the Reynolds stress divergence over the whole domain, as discussed at the end of section \ref{sec:ZSWM}. According to equation (\ref{eq:BC_RSB}),  the amplitude of the boundary forcing evaluated at the bottom scales as $\epsilon ^2 Re^{1/2} /F\!r$. In the limit $ReF\!r^2\gg 1$, this amplitude is much larger than in the free\--slip case. In addition, it increases with the Reynolds number. However, we will see in section \ref{sec:BoundaryFlows} that the amplitude does not blow up in a distinguished limit that is consistent with the linearization of the equations.}

\subsection{Boundary flows\label{sec:BoundaryFlows}}

{We now  look for the mean flow response to the Reynolds stress divergences, by inserting the linear predictions for  wave fields into equation (\ref{eq:Evol_MeanU}).  {When ignoring the influence of the mean flow on the wave fields, equation (\ref{eq:Evol_MeanU}) becomes a linear diffusion equation with a steady forcing, that can be decomposed into a bulk and a boundary contribution, as in equation (\ref{eq:BulkBL_decomp}).}

 { The typical time scales $\tau_{w}$ and $\tau_{bl}$ for the mean flow to reach a given velocity $U$ in the presence of either bulk or boundary streaming forcing terms are obtained by balancing $\p_{t}\overline{u}$ with $\p_{z}F_{w}$ and $\p_{z}F_{bl}$, respectively. Using the large Reynolds asymptotic  estimates given in equation (\ref{eq:BC_RSB}) leads then to $\tau_{bl}/\tau_{w}\sim 1/(F\!r^{2}Re)$ in the free-slip case and $\tau_{bl}/\tau_{w}\sim 1/(F\!r^{2}Re)^{3/2}$ in the no-slip case. We thus expect the boundary streaming to dominate over the bulk streaming at the early stage of the mean flow evolution in both cases.} 

 {At a quasi\--linear level, the early stage of the mean flow evolution is obtained for both the free\--slip and the no\--slip conditions by solving equation (\ref{eq:Evol_MeanU}) numerically, assuming that the wave field is described by equations (\ref{eq:DispersionRel}), (\ref{eq:WaveGeneral}), (\ref{eq:PolarWave}) and (\ref{eq:BC_Fi}).} A finite size domain is considered in the simulations with  {an aspect ratio} $r=6L_{Re}$.  
The waves are computed as if the domain were semi-infinite and a free-slip upper boundary condition is considered for the mean flow.} 
\begin{figure}
	\centering \includegraphics[width=1.\linewidth]{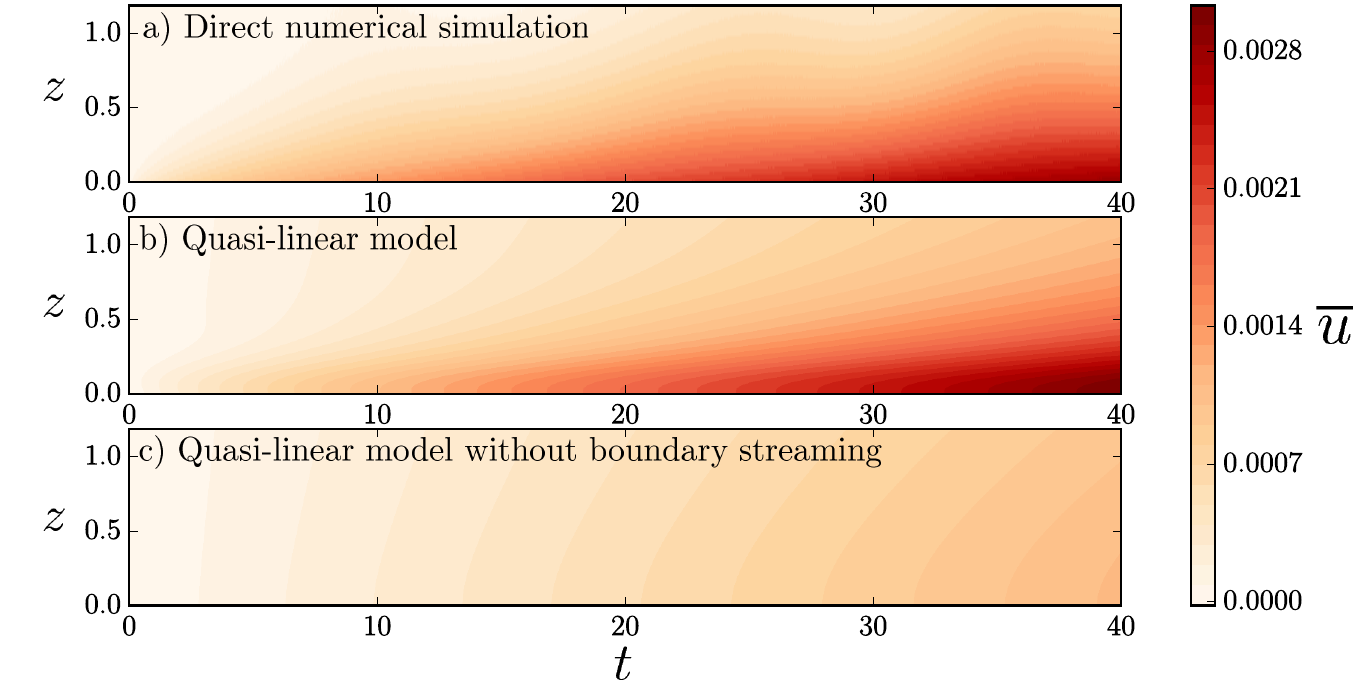}
	\caption{\label{fig:Hov_MIT} Hovm\"oller diagrams of the mean flow, $\overline{u}\left(z,t\right)$,  {for the scenario in which the bottom boundary condition is free-slip.}  a) Direct numerical simulation (DNS)  b)  quasi\--linear model  c) quasi\--linear model  without the boundary streaming terms in the Reynolds stress divergence. The parameters are $Re=200$, $F\!r=0.3$ and $\epsilon=0.01$, $\mathrm{d}x=\mathrm{d}z=\delta_{{Re}}/15$. The grid is exponentially stretched on the vertical axis above $z=6L_{Re}$ in the DNS.  { At larger time, around $t \sim 300$, the mean flow induced by bulk streaming becomes larger than the mean flow induced by boundary streaming}.}
\end{figure}

In figure \ref{fig:Hov_MIT}, we compare the quasi\--linear predictions for the free\--slip boundary condition against direct numerical simulations. The parameters are $Re=200$, $F\!r=0.3$ and $\epsilon=0.01$. For those parameters, the wave boundary layer thickness is $\delta_{   {Re}}=0.1$ and the viscous damping length is $L_{   {Re}}=5.15$. The Hovm\"oller diagrams focus on an area close to the bottom boundary.  We use a vertical resolution of $\mathrm{d}z=0.0067$ which resolves properly the wave boundary layer.  {In the DNS, a stretched grid has been implemented} on the vertical to avoid any downward reflection. The quasi\--linear model captures well the boundary streaming effect. To emphasise the crucial role of the boundary streaming term, we added a diagram {in} figure \ref{fig:Hov_MIT} of a quasi\--linear computation where the boundary forcing has been removed in (\ref{eq:Evol_MeanU}) ( {$F_{bl}=0$} in (\ref{eq:BulkBL_decomp})). We clearly see that the presence of boundary streaming is important to predict accurately  {the early evolution} of the mean flow in this case. 

In figure \ref{fig:LAB}-c, we show a Hovm\"oller diagram of the mean flow computed using the quasi\--linear model in the case of no\--slip boundary condition. The parameters are $Re=200$, $F\!r=0.3$ and {$\epsilon=0.005$}.  {As expected from the discussion following equation (\ref{eq:BC_RSB}), the boundary forcing generates a strong boundary mean flow going in  a direction opposite to the direction of the bulk mean flow. Consistently with our previous estimates of typical timescales for the mean flow evolution,  the establishment of the bulk flow occurs at a time scale larger than the establishment of the quasi\--stationary boundary flow. }

 {In the no-slip case, the mean flow eventually reaches a stationary state given by 
\begin{equation}
\overline{u}_{\infty}\left(z\right)=Re\int_{0}^{z}\overline{u^{\prime}w^{\prime}}\left(z^{\prime}\right)\mathrm{d}z^{\prime}.\label{eq:Stationary_NS}
\end{equation} 
Then, the contribution from the boundary streaming is negligible with respect to the contribution from the bulk streaming. This can be quantified by computing the order of magnitude of typical mean flow amplitudes $U_{w}$ and $U_{bl}$ obtained by splitting Reynolds stresses in (\ref{eq:Stationary_NS}) into a bulk and a boundary contribution, respectively. Using the large Reynolds asymptotic expressions obtained in (\ref{eq:BC_RS}) assuming $ReF\!r^2 \rightarrow +\infty$ and $F\!r\rightarrow 0$, we get $U_{w}\sim ( \epsilon Re F\!r)^2$ and $U_{bl}\sim\epsilon^{2}Re^{1/2}F\!r^{-1}$. Their ratio scale as $(Re F\!r^2)^{3/2} $, and thus tends diverges: the bulk flow is dominant in the longtime limit.} 

 In the free-slip case, no stationary regime is reached and the mean flow amplitude keeps increasing in time. It can be assessed by considering the $z$-integrated momentum, $P\left(t\right)=\int_{0}^{\infty}\overline{u}\left(z,t\right)\mathrm{d}z$. Using the free-slip boundary condition and integrating (\ref{eq:Evol_MeanU}), we get $P\left(t\right)=\left(\overline{u^{\prime}w^{\prime}}|_{z=0}\right)t$. At sufficiently large times, the mean flow varies over the  {characteristic length scale $\sqrt{t/Re}$}. Consequently, the mean flow amplitude $ P/ \sqrt{t/Re}$ increases as $t^{1/2}$: eventually, the feedback of the mean flow on the wave will no longer be negligible. We can however use this mean flow amplitude estimate, together with the large Reynolds asymptotic expressions in (\ref{eq:BC_RS}), to infer $U_{bl}/U_{w}\sim 1/(F\!r Re^{1/2})$.  {This scaling has been obtained under the assumption $F\!r^2 Re\rightarrow +\infty$. This means that the bulk flow is dominant in the long time limit, just as in the no\--slip case. It is also possible to estimate the time scale $\tau$ for which the mean flow induced by the bulk streaming becomes of the same order as the mean flow induced by the boundary streaming. When this occurs, the long time limit is relevant for the estimate of the mean flow induced by boundary streaming, as above: $U_{bl}  \sqrt{\tau /Re} \sim  F_{bl} (0)$. By contrast, assuming $L_{Re} \gg \delta_{Re}$, the flow induced by the bulk streaming must be estimated using an early time limit:  $U_w \sim \tau \partial_z F_w|_{z=0}$. Then, using $U_{bl} \sim U_w$ yields $\tau \sim F\!r^4 Re^{2}$. Using the parameters corresponding to figure 4 yields $\tau \sim 300$.}

\subsection{Limitation of the quasi\--linear model}

\begin{figure}
	\begin{center}
	\includegraphics[width=0.6\linewidth]{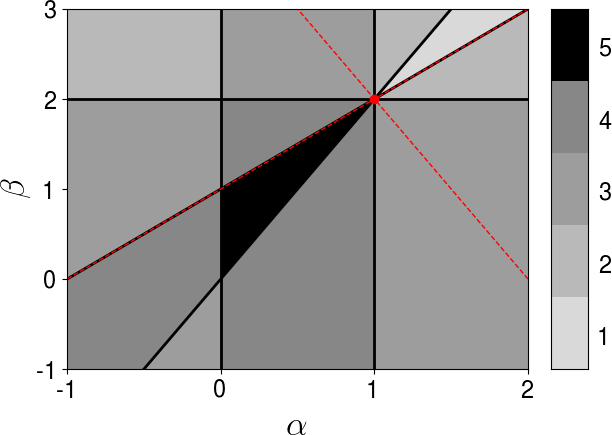}
	\caption{\label{fig:distinguished} {Distinguished limit for the validity of the linear dynamics around a state of rest in the no\--slip case: $(F\!r,Re)=(\epsilon^{\alpha},\epsilon^{-\beta})$.  Each line delimitates a half plane where one of the constraints is satisfied. The colourmap shows the number of constraints that are satisfied. The black region corresponds to the range of exponents $(\alpha,\beta)$ for which the asymptotic approach is self-consistent: all the constraints are satisfied for those scalings. In the free\--slip case, there is an additional constraint $\beta<4\alpha-2$ which is not represented here, but that is fulfilled within the black area. The dash red lines correspond to limit cases above which the mean flows induced by the bulk streaming and boundary streaming impact the wave field. The red dot corresponds to the regime $(\alpha,\beta)=(1,2)$ where the distinguished limit is marginally satisfied, and where two-way coupling between waves and mean flow can no longer be neglected.}}
		\end{center}
\end{figure}

 {To derive the quasi\--linear model around a state of rest presented above, the only necessary assumption is $\epsilon \rightarrow 0$, with all other parameters fixed. 
The quasi\--linear numerical calculations have been made using the actual solution of the dispersion relation (\ref{eq:DispersionRel}), but we obtained  scalings by assuming simplified expressions for the wave field in the inviscid limit $F\!r^2 Re\rightarrow +\infty $, together with the hydrostatic limit $F\!r \rightarrow 0$. These two conditions imply $\delta_{Re}/L_{Re}\sim \left( Re F\!r^2 \right)^{-3/2}\rightarrow 0$, and therefore make possible a clear distinction between a bulk and a boundary contribution to streaming. To establish a self-consistent distinguished limit, we write 
\begin{equation}
(\epsilon,F\!r,Re)=(\epsilon,\epsilon^{\alpha},\epsilon^{-\beta}).
\end{equation}
The two simplifying assumptions above correspond to $\beta>2\alpha$ and $\alpha>0$. Let us now list the conditions required for the validity of the linearisation procedure. 
In the bulk, neglecting the nonlinear (advection) terms with respect to the viscous terms and the time derivative terms yields to the conditions  $ \beta>1+\alpha$ 
and $\alpha<1$, respectively. Similarly, neglecting the nonlinear terms with respect to the viscous term in the boundary layer yields to $\beta<4\alpha-2$ 
in the free-slip case, and to $\alpha<1$ in the no-slip case. Finally, expressing the boundary condition at $z=0$ instead of $z=h$ for both free-slip and no-slip requires $\beta<2$. 
This condition also guarantees the validity of neglecting nonlinear terms in the bottom boundary conditions (\ref{eq:BC}). There are therefore six inequalities to be satisfied for  $\alpha$ and $\beta$ in the free-slip case, five inequalities for the no-slip case.  For the latter case, regimes of parameters for which these conditions are all fulfilled in presented in figure \ref{fig:distinguished}. The black area corresponds to regimes fulfilling all the constraints. The additional condition required for the free-slip case is also fulfilled within this black area.}

 {In all the above analysis, we have neglected the feedback of the mean flow on the wave field. This is always valid at sufficiently short times. However, we saw that this can never be satisfied at large time in the free-slip case since the mean flow keeps increasing in time. In the no\--slip case, we found that both the bulk and the boundary mean flow are indeed negligible with respect to the horizontal wave velocity, as $U_{w}\sim ( \epsilon Re F\!r)^2 \rightarrow 0$ and   $U_{bl}\sim \epsilon^{2}Re^{1/2}F\!r^{-1} \rightarrow 0$ in the distinguished limit.}

 {In the no\--slip case, we expect a two-way coupling between waves and mean flow when the induced flows are of order one \-- with $\epsilon Re F\!r\sim  1$ for the bulk flow and $\epsilon Re^{1/4}F\!r^{-1}\sim 1$ \-- since the terms involving the mean flow can no longer be ignored to compute the wave field in that case. These additional conditions are represented by the dashed red lines in figure \ref{fig:distinguished}. The red dot corresponds to the regime $(\alpha,\beta)=(1,2)$ satisfying marginally the distinguished limit while  allowing for order one mean flows induced by both the bulk and the boundary forcing. Within this regime the viscous terms are of the order of nonlinear terms in the bulk wave equation, thus invalidating the quasi\--linear approach. This limitation can be bypassed by introducing additional dissipative terms, such as a linear friction term in the zonal flow equation or the buoyancy equation. Such terms would allow us to control the typical vertical length scale for wave attenuation, related to the intensity of the bulk internal wave streaming, without varying the Reynolds number that constrains the mean flow vertical gradients. To avoid the introduction of such additional parameters, we choose in the following to consider the quasi\--linear equations as an \textit{ad hoc} model for wave-mean flow interactions. This simplified model will illustrate how boundary streaming can affect mean flow properties in the bulk when the feedback of the mean flow on the wave field is taken into account.}

\section{{Application to an idealised analogue of the quasi\--biennial oscillation}}\label{sec:MeanFlow}

{We consider  a standing wave pattern imposed at the bottom boundary: $h\left(x,t\right)=\cos\left(x\right)\cos\left(t\right)$ with a no\--slip boundary condition. This idealised configuration is thought to capture the essential mechanism at the origin of the equatorial stratospheric quasi\--biennial oscillation \citep{Plumb1977}, and has been experimentally studied by \citet{Plumb1978}. Two linear waves with equal amplitude and opposite zonal phase speeds are emitted by such a bottom excitation. The resulting Reynolds stress is simply the sum of the Reynolds stresses computed from each individual wave plus a rapidly oscillating term that can be smoothed out by averaging over this fast oscillation. The Reynolds stress divergences induced by the two waves are opposite and anneal each other in the absence of mean flow.  {Above a certain value of the amplitude of the waves, a Hopf bifurcation occurs: a vacillating mean flow is generated and approaches }a limit cycle \citep{Plumb1977}.  \cite{Plumb1978} reported the spontaneous generation of an oscillating  mean flow in laboratory experiments when the wave amplitude exceeds a threshold, and compared their measurements against quasi\--linear computations. They considered a no\--slip bottom boundary condition for the mean flow but inviscid impermeability condition  for the wave field, allowing them to ignore any boundary layer effect. Here, we investigate the effect of the viscous boundary layers and the associated boundary streaming on the oscillation arising with the standing wave excitation, assuming a no\--slip condition for both the mean flow and the waves. We show that  {the inclusion of boundary streaming induces important alterations on the mean flow} in this idealised model of wave-mean flow interactions.}

{In section \ref{sec:ViscIW}, we ignored the effect of the mean flow on the wave field. We need here to take this feedback into account, as the initial instability arises from a perturbation of the  mean flow itself. The effect of the mean flow on the wave is included by performing a WKB expansion of the wave field following the method of \citet{Muraschko2015}, but including dissipative effects. The full calculation is detailed in appendix \ref{sec:WKB}. The Reynolds stress divergence is then computed and inserted  into the mean flow equation (\ref{eq:Evol_MeanU}) in order to compute the long-time evolution of $\overline{u}$. This task is done numerically using the results of appendix \ref{sec:WKB} and the no-slip boundary condition in (\ref{eq:BC_Fi}).  {While \citet{Plumb1978} considered asymptotic expression for the bulk solution of the dispersion relation  (\ref{eq:WKBDisp}), our numerical calculations use the actual solutions. As discussed at the end of Appendix A, this solution captures important corrections close to the critical layers, where the mean flow is of the order of the wave zonal phase speed.}

\begin{figure}
	\begin{center}
		\subfloat[$Re=67$, $L_{Re}=0.44$ and $\delta_{Re}=0.13$]{\includegraphics[width=0.658\linewidth]{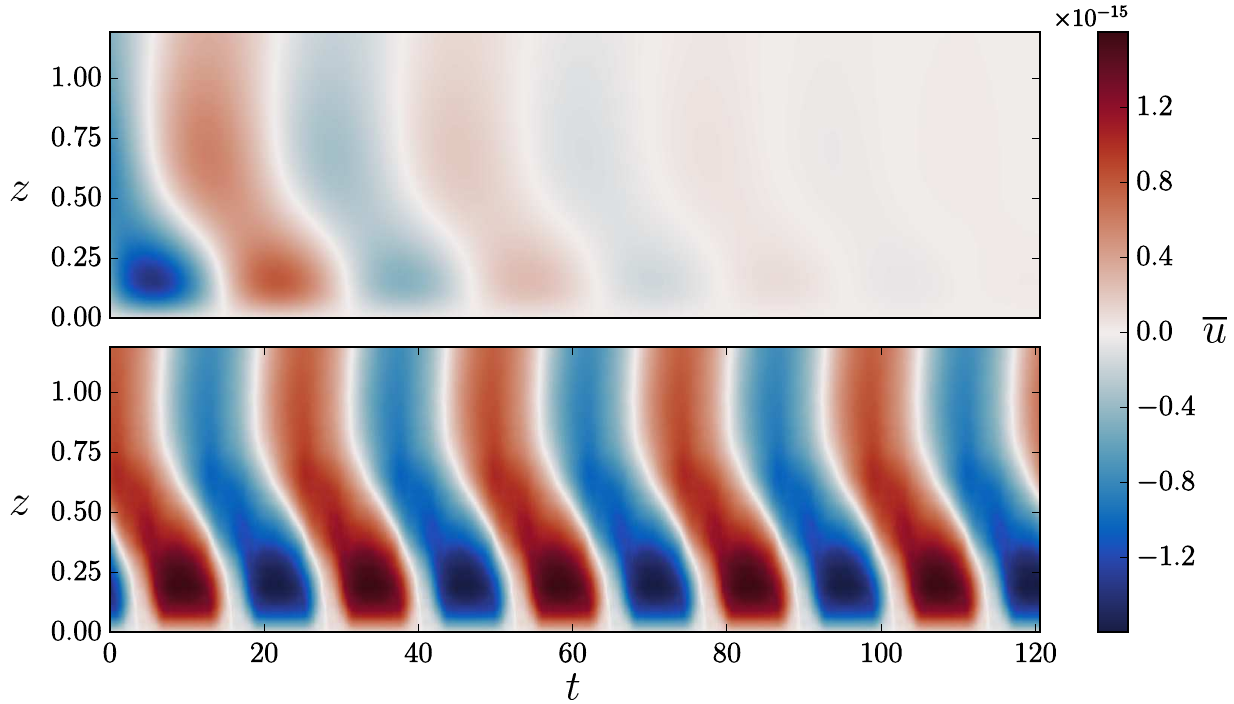}\label{sub:QBO1}}\\
		\subfloat[$Re=154$, $L_{Re}=0.65$ and $\delta_{Re}=0.10$]{\includegraphics[width=0.658\linewidth]{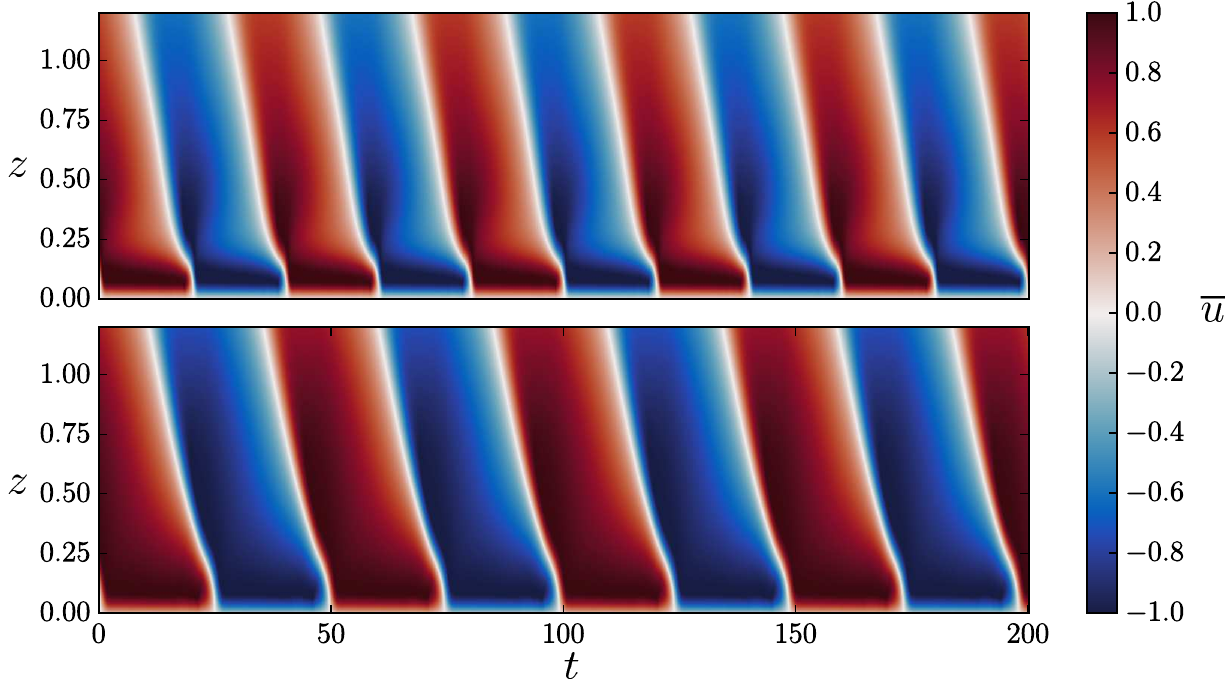}\label{sub:QBO2}}\\
		\subfloat[$Re=200$, $L_{Re}=0.78$ and $\delta_{Re}=0.089$]{\includegraphics[width=0.658\linewidth]{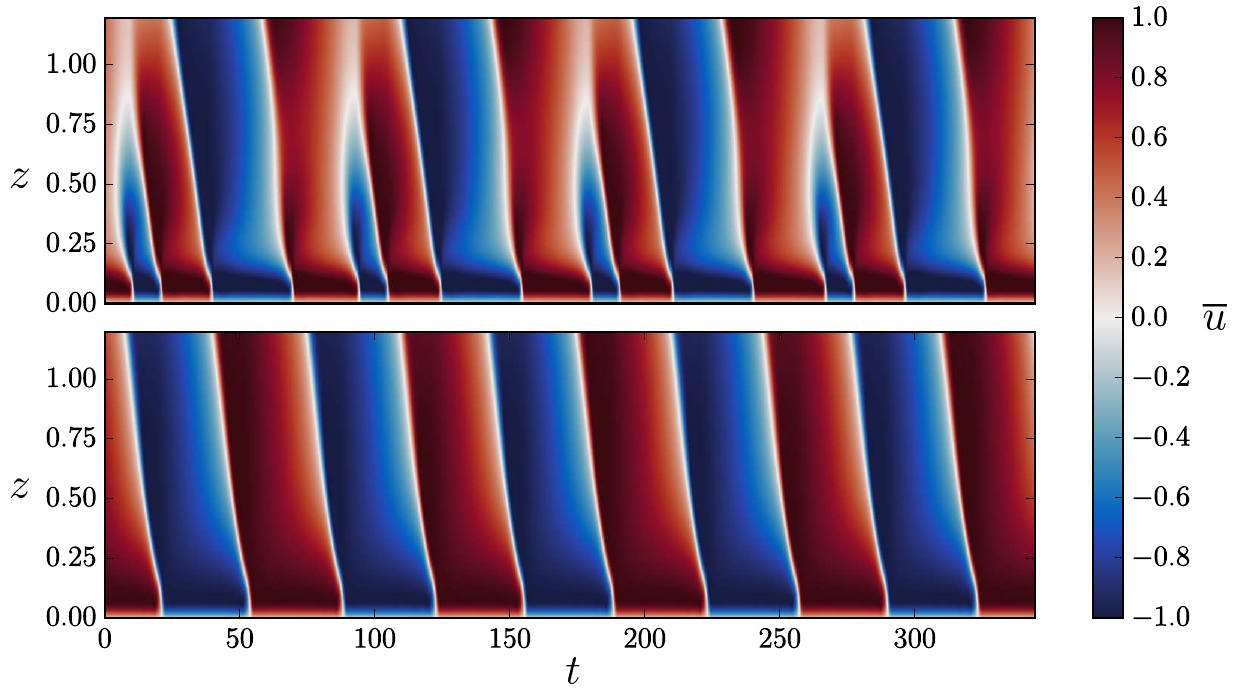}\label{sub:QBO3}}
	\caption{\label{fig:Qbo} {The mean flow, $\overline{u}\left(z,t\right)$, is generated by the streaming coming from two counter propagative waves with the same amplitude and opposite horizontal phase speed, generated by a vertically oscillating bottom boundary with no\--slip condition using the quasi\--linear model. Hovm\"oller diagrams of the mean flow time series are shown for three different Reynolds numbers. In each panel (a,b,c) the upper plot corresponds to a case where the boundary streaming has been included in the computation while the lower plot corresponds to a computation with same parameters but without the contribution of the boundary streaming terms. In all cases, $F\!r=0.15$ and $\epsilon=0.3$. In figure b), the mean flow oscillates with an oscillation period of about $40$ and $50$ time units for the case with (upper plot) and without (lower plot) boundary streaming respectively.}}
	\end{center}
\end{figure}

The resulting  {Hovm\"oller diagrams} of mean flows time series are showed on figure \ref{fig:Qbo}  {for different values of the Reynolds number}. The time series used for the upper plots have been computed using the full quasi\--linear model while the one used for the bottom plots have been computed without the boundary layer contributions.  All simulations start with the same initial perturbation. In figure \ref{sub:QBO1}, we see that the inclusion of boundary streaming  {has altered the critical parameter values at which the bifurcation to mean-flow reversals occurs}. In figure \ref{sub:QBO2}, the Reynolds number is increased and the oscillation is present in both cases. However, the oscillation period is decreased by 20\% when the boundary streaming is included. By further increasing the Reynolds number we see in figure \ref{sub:QBO3} that the inclusion of the boundary streaming significantly changes the mean flow oscillation. This new regime presenting an additional frequency in the signal can actually be reached without the boundary streaming but at a larger Reynolds number. A similar regime has  been reported by \cite{MacGregor2001}, and we will more thoroughly study these bifurcations in a companion paper. Our aim is here to show that the presence of the wave boundary layer has an impact on such bifurcation diagrams, in addition to significantly altering the period of oscillations in the periodic case.}
\begin{figure}
	\centering \includegraphics[width=\linewidth]{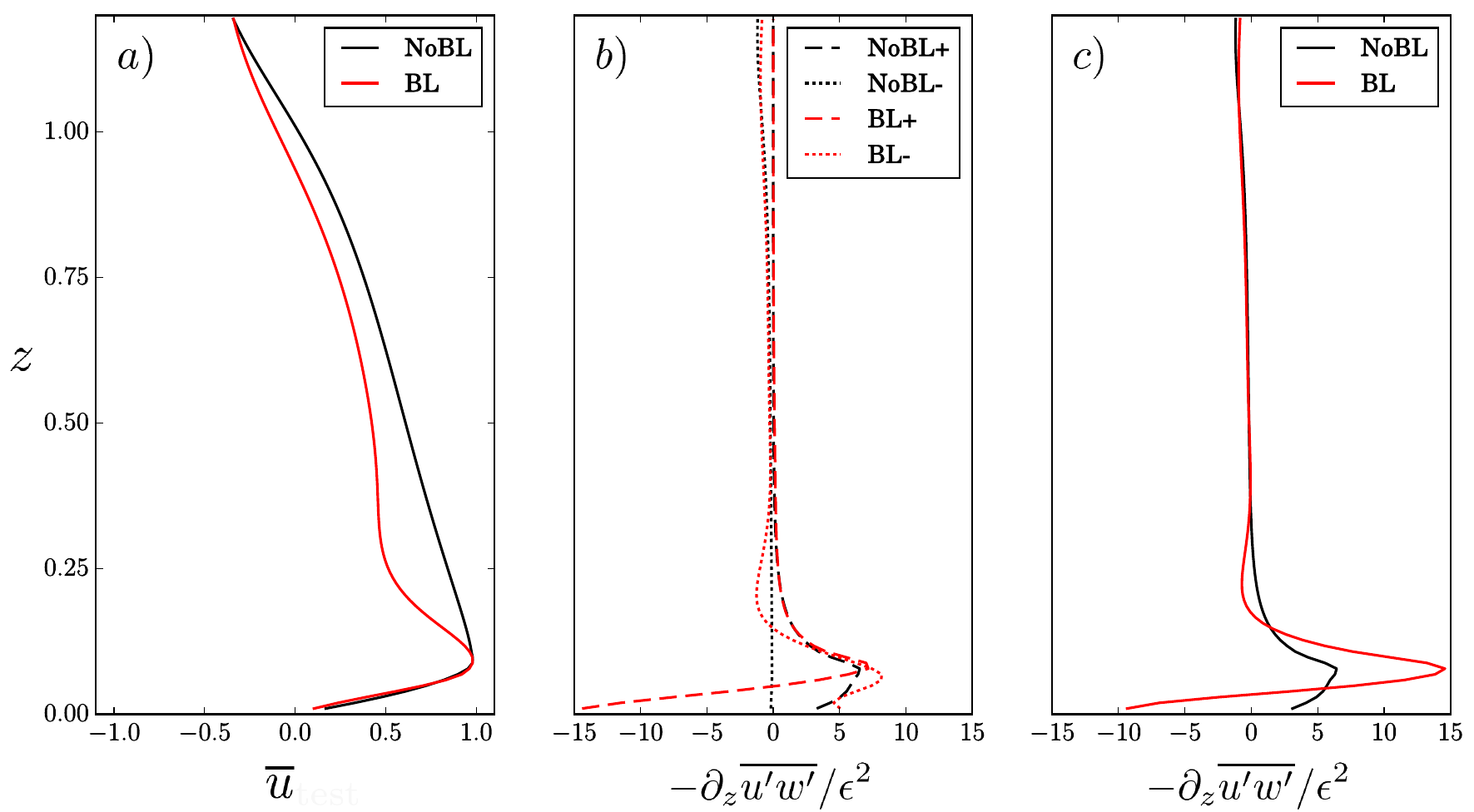}
	\caption{\label{fig:Stream_QBO}  { $a)$ Mean flow snapshots extracted from the two time series plotted in figure \ref{sub:QBO2} computed with (BL) and without (NoBL) boundary streaming and taken at $t=10$ and $t=12.5$ respectively. $b)$ Plot of the associated Reynolds stress divergences obtained for the wave propagating in the direction of positive mean flow (``+'') and for the counter\--propagating one (``-'') considering the mean flow profile obtained by either including (BL) or ignoring (NoBL) the boundary streaming. $c)$ Plot of the total Reynold stress divergences, the sum of the two counter propagative waves contributions for the case with boundary streaming (BL) and without boundary streaming (NoBL).}}
\end{figure}

 For the range of parameters corresponding to figures \ref{sub:QBO2} and \ref{sub:QBO3}, the mean flow reaches an amplitude close to the phase speed of the waves. To investigate the effect of the boundary layers in these cases, we consider two mean flow snapshots, plotted in figure \ref{fig:Stream_QBO}, taken from the two time-series plotted in figure \ref{sub:QBO2}. The snapshots are taken at the same stage of the oscillation cycle.

 {The Reynold stress divergences computed using the two mean flow snapshots shown in figure \ref{fig:Stream_QBO}-a  and considering the counter\--propagating waves separately are plotted in figure \ref{fig:Stream_QBO}-b. The total Reynolds stress divergences are plotted in figure \ref{fig:Stream_QBO}-c. As expected, the boundary layers significantly modify the streaming vertical profiles. Interestingly, while bulk streaming is dominated by the wave travelling in the same direction as the mean flow, the main discrepancy between the case with and without boundary streaming comes from the boundary forcing associated with the wave going in a direction opposite to the mean flow at the bottom.}
 
   In figure \ref{sub:QBO2}, we see that the bottom profile of the mean flow is approximately steady before a reversal.  Let us call $\lambda_{Re}$ the typical length  over which the mean flow reaches its extremum value. This velocity is of order one as it is close to the gravity wave zonal phase speed.  {Using equation (\ref{eq:Evol_MeanU}), we  infer the typical scaling of $\lambda_{Re}$ by balancing the viscous term $Re^{-1}\p_{zz}\overline{u}$ and the streaming term $\p_{z}\overline{u^{\prime}w^{\prime}}$. This last term is dominated by the bulk contribution $F_w$ associated with the wave that propagates in the same direction as the mean flow. This yields $\lambda_{Re}\sim \left(Re\, F_w(0) \right)^{-1}$. The order of magnitude of $F_w(0)$ can be obtained by using the asymptotic expression in equation (\ref{eq:BC_RS}), under the assumption $F\!r\rightarrow 0$ and $Re^2 F\!r \rightarrow \infty $. This yields  $\lambda_{Re} \sim F\!r / (\epsilon^{2}Re)$.  Using parameters of figure \ref{fig:Stream_QBO}, we find  $\lambda_{Re}\sim 0.1$.}
 
  { We expect boundary streaming to have a significant impact on mean flow reversals when such reversals occur within the boundary layer, i.e. when $\lambda_{Re} \sim  \delta_{Re}$ or $\lambda_{Re} \ll  \delta_{Re}$. This is indeed the case in figure \ref{sub:QBO3}, where $\delta_{Re}\approx 0.1$. It is instructive to establish the range of parameter for which this condition is satisfied.  Using   $\delta_{Re} \sim 1/Re^{1/2}$, we find that this length scale is larger or of the order of $\lambda_{Re}$ when $F\!r /(\epsilon^2Re^{1/2})\rightarrow 0$. The above analysis suggests the possibility for active control of the boundary layer on the bulk flow when  $ (F\!r,Re)=(\epsilon^{\alpha}, \epsilon^{-\beta})$ with $\beta \ge 4-2\alpha$, $\alpha>0$ and $\beta>2\alpha$.}
 
 {As seen on figure  \ref{fig:distinguished}, the distinguished limits consistent with an active control of the boundary layer on bulk flow reversal in the \textit{ad hoc} quasi\--linear model do not overlap with the distinguished limits ensuring the validity of the quasi\--linear dynamics around a state of rest. In fact, both sets of constraints can only be marginally satisfied at the point  $(\alpha,\beta)=(1,2)$. This is the main caveat of the analysis presented in this section and illustrated in figure \ref{fig:Stream_QBO}: because $\epsilon Re F\!r\sim 1$, the nonlinear terms involving bulk waves are not negligible with respect to the bulk viscous term. Furthermore, since $\epsilon Re^{1/2}\sim 1$, the boundary layer can hardly be considered as linear. Whether active control of the boundary on the bulk flow persists when nonlinear effects are added back into the problem will need to be addressed in a future work. }

{ Another caveat of the quasi\--linear model presented here  comes from the assumptions underlying the WKB approach used to compute the wave-field. We assumed that there is a vertical scale separation between the wave and the mean flow and that the wave field reaches its steady state in a time much shorter than the typical time of evolution of the mean flow. The wave field is thus computed using a static WKB approximation with a frozen in time mean flow. Since the mean flows shown in figure \ref{fig:Qbo} exhibit sharp shear at the bottom, and since they  {reach values of the order of the zonal phase speed of the waves}, those hypotheses are not valid. Nevertheless, this WKB approximation is the simplest way of accounting for the mean flow effect on the wave field, and a useful first step to understand their interactions.



We should finally stress that the no-slip bottom boundary condition is most certainly irrelevant to the actual quasi\--biennial oscillation occurring in the upper atmosphere and that our model has been derived in a distinguished limit for which the viscous boundary layer is much larger than the boundary height variations, which is not satisfied in laboratory experiments. However, despite these limitations, our results show that the boundary conditions and the associated wave boundary layers should not be overlooked, since boundary streaming may have a quantitative impact on mean flow reversals in the domain bulk.}

\section{Conclusion}\label{sec:conclusion}

 {We have shown that changing the boundary conditions has a  significant impact on the boundary mean flow generated by internal waves emitted from an undulating wall in a viscous stratified fluid.
We first compared the effect of no\--slip and free\--slip boundary conditions  by considering a distinguished limit that makes possible a clear separation between the bulk and a viscous boundary layer. 
In the no-slip scenario, the Reynolds stress  {divergence} scales at early time in direct proportion to $\epsilon^{2}\sqrt{Re} / F\!r $, where $\epsilon$ is the dimensionless wave amplitude, $Re$ is the wave Reynolds number, and $F\!r$ the wave Froude number. However, bulk streaming dominates over boundary streaming in the large time limit, and the system reaches a stationary state with a mean flow that remains negligible with respect to the wave-field. In the free\--slip scenario, boundary forcing amplitude does not depend on the Reynolds number, only its e-folding depth does. The presence of the boundary layer qualitatively alters the early time flow evolution. Just as in the no-slip case, bulk streaming has a dominant contribution at large\--time. However, contrary to the no-slip case, the system does not reach a stationary state. In both cases, the distinguished limit considered to derive these results prevent a two-way coupling between waves and mean flows.}

 {To address the interplay between boundary streaming, waves and mean-flows, we treated the case of a forced standing wave with an ad hoc truncation of the dynamics based on a quasi\--linear approach. This model captures the basic mechanism responsible for the quasi\--biennial oscillation \citep{Plumb1977}.   Using a novel WKB treatment of the waves that takes into account viscous effects, we investigated the effect of boundary streaming on mean flow reversals in this model. We found that boundary streaming significantly alters the mean flow reversals by either inhibiting them, decreasing their period or altering their periodicity depending on the wave amplitude. Further work will be needed to determine whether this active control of bulk properties by boundary streaming is robust to the presence of wave-wave interactions.} 

Beyond these particular examples, our results show the importance of describing properly the physical processes taking place in the boundary layers where waves are emitted to model correctly the large-scale flows induced by these waves. 
 We have neglected the effects of rotation and diffusion of buoyancy which are known to change the properties of the wave fields close to boundaries \citep{Grisouard2015,Grisouard2016}, and will, therefore, affect boundary streaming. By restricting ourselves to a quasi\--linear approach, we have also neglected nonlinear effects that may become important close to the boundary, even in the limit of weak undulations, due to the emergence of strong boundary currents. All these effects could deserve special  attention is future numerical and laboratory experiments. \\ 

The authors warmly thank Louis\--Philippe Nadeau for his help with the MIT-GCM, and express their gratitude to Freddy Bouchet and Thierry Dauxois for their useful insights.

\appendix
\section{WKB expansion of viscous internal gravity wave within a frozen in time mean flow}\label{sec:WKB}
We compute here the leading order terms of a Wentzel\--Kramers\--Brillouin (WKB) expansion of the viscous wave field within a weakly sheared mean flow frozen in time. We follow the method developed in \citet{Muraschko2015}, the novelty being the presence of viscosity, in the wave equation (\ref{eq:GlobalFluct}). {The internal wave equations write
\begin{equation}\label{eq:LinWithMean}
	\left\{
	\begin{array}{ll}
		\p_{t}u^{\prime}+\overline{u}\p_{x}u^{\prime}+w^{\prime}\p_{z}\overline{u}+\p_{x}p^{\prime}-Re^{-1}\nabla^{2}u^{\prime}&=0\\
		\p_{t}w^{\prime}+\overline{u}\p_{x}w^{\prime}+\p_{z}p^{\prime}-F\!r^{-2}b^{\prime}-Re^{-1}\nabla^{2}w^{\prime}&=0\\
		\p_{t}b^{\prime}+\overline{u}\p_{x}b^{\prime}+w^{\prime}&=0\\
		\p_{x}u^{\prime}+\p_{z}w^{\prime}&=0
	\end{array}
	\right..
\end{equation}
We assume that the mean flow is time independent and varies over a vertical scale $L_{u}$ much larger than the inverse of the vertical wave number modulus $1/\left|m\right|$. None of those quantities is known prior to our problem. For the present calculation, we assume $L_{u}\gg1$ and $\left|m\right|\sim 1$ but the final result will apply for different scalings as long as $L_{u}m\gg1$ is fulfilled. We therefore assume that $\overline{u}$ depends on a smooth variable $Z=az$ with $a=1/ L_{u} \ll1$.} 

We introduce the WKB ansatz for a monochromatic wave solution 
\begin{equation}\label{eq:WKB_Exp}
	\left[
	\begin{array}{c}
		u^{\prime}\\
		w^{\prime}\\
		b^{\prime}\\
		p^{\prime}
	\end{array}
	\right] = \Re\left\{\sum_{j=0}^{+\infty}a^{j} \left[
	\begin{array}{c}
		\tilde{u}_{j}\left(Z\right)\\
		\tilde{w}_{j}\left(Z\right)\\
		\tilde{b}_{j}\left(Z\right)\\
		\tilde{p}_{j}\left(Z\right)
	\end{array}
	\right]e^{\displaystyle i{\left(x-ct\right)}+\frac{i\Phi\left(Z\right)}{a}}\right\}    {,}
\end{equation}
with $c=\pm1$. The function $\Phi\left(Z\right)$ accounts for the vertical phase progression of the wave. The local vertical wave number is defined by $m\left({Z}\right)=\p_{Z}\Phi$. Inserting this expansion into the previous equation and collecting the leading order terms in $a$ leads to:
\begin{equation}\label{eq:WKBSolv}
	\mathsfbi{M} \left[
	\begin{array}{c}
		\tilde{u}_{0}\\
		\tilde{w}_{0}\\
		\tilde{b}_{0}\\
		\tilde{p}_{0}
	\end{array}
	\right]+a\left(\mathsfbi{M}\left[
	\begin{array}{c}
		\tilde{u}_{1}\\
		\tilde{w}_{1}\\
		\tilde{b}_{1}\\
		\tilde{p}_{1}
	\end{array}
	\right]+\left[
	\begin{array}{c}
		\tilde{w}_{0}\p_{Z}\overline{u}-i{Re^{-1}}\left(\tilde{u}_{0}\p_{Z}m+2m\p_{Z}\tilde{u}_{0}\right)\\
		\p_{Z}\tilde{p}_{0}-i{Re^{-1}}\left(\tilde{w}_{0}\p_{Z}m+2m\p_{Z}\tilde{w}_{0}\right)\\
		0\\
		\p_{Z}\tilde{w}_{0}
	\end{array}
	\right]\right)+o\left(a\right)=0   {,}
\end{equation}
with
\begin{equation}
	\mathsfbi{M}=
	\left[
	\begin{array}{cccc}
		{Re^{-1}\left(1+m^{2}\right)-i\left(   {c}-\overline{u}\right)} & 0 & 0 & {i}\\
		0 & {Re^{-1}\left(1+m^{2}\right)-i\left(   {c}-\overline{u}\right)} & -1 & im\\
		0 & {F\!r^{-2}} & -i{\left(   {c}-\overline{u}\right)} & 0\\
		{i} & im & 0 & 0
	\end{array}\right].
\end{equation}

We introduce the polarisation $\mathbf{P}\left[m\right]$ defined by $\left[\tilde{u}_{0},\tilde{w}_{0},\tilde{b}_{0},\tilde{p}_{0}\right]=\phi_{0}\left(Z\right)\mathbf{P}\left[m\right]$, where $\phi_{0}\left(Z\right)$ is the amplitude of the wave mode. The cancellation of the zeroth order term in equation (\ref{eq:WKB_Exp}) yields to $\det\mathsfbi{M}=0$. This gives the local dispersion relation 
\begin{equation}{
	F\!r^{2}\left(c-\overline{u}\right)^{2}\left(1+m^{2}\right)\left(1+iRe^{-1}\frac{1+m^{2}}{c-\overline{u}}\right)=1} \ .\label{eq:WKBDisp}
\end{equation} 
Then, using $\mathsfbi{M}\,\mathbf{P}=0$ we obtain the polarisation expression 
\begin{equation}{
	\mathbf{P}\left[m\right]=\left[c-\overline{u},-\frac{1}{m}\left(c-\overline{u}\right),\frac{i}{F\!r^{2}m},\frac{1}{F\!r^{2}\left(1+m^{2}\right)}\right]}.
\end{equation}

The cancellation of the terms proportional to $a$ in (\ref{eq:WKBSolv}) provides an equation for the amplitude $\phi_{0}\left(Z\right)$. To get rid of the terms involving components of the order one wave, {let us introduce the vector $\mathbf{Q}=\left[\tilde{u}_{0},\tilde{w}_{0},-\tilde{b}_{0}F\!r,\tilde{p}_{0}\right]$ such that $\mathbf{Q}^{\bot}\,\mathsfbi{M}=\boldsymbol{0}$. We then take the inner product between $\mathbf{Q}$ and the terms proportional to $a$ in (\ref{eq:WKBSolv}). This gives
\begin{equation}
	\tilde{u}_{0}\tilde{w}_{0}\p_{Z}\overline{u}+\p_{Z}\left(\tilde{w}_{0}\tilde{p}_{0}\right)=iRe^{-1}\p_{Z}\left(m\left(\tilde{u}_{0}^{2}+\tilde{w}_{0}^{2}\right)\right).
\end{equation}
By introducing $\varphi_{0}^{2}=\phi_{0}^{2}\left(c-\overline{u}\right)^{2}/m$ and using the dispersion relation (\ref{eq:WKBDisp}), we obtain after some algebra:
\begin{equation}
	\p_{Z}\log{\varphi_{0}^{2}}+\frac{2iRe^{-1}\left(1+m^2\right)}{c-\overline{u}+2iRe^{-1}\left(1+m^2\right)}\p_{Z}\log\left(1+m^{2}\right)=0   {.}
\end{equation}
This last equation has to be solved for every solution $m\left(Z\right)$ of the dispersion relation. This is done numerically in general. 

 {By solving the dispersion relation (\ref{eq:WKBDisp}), we find that in the limit of large Reynolds number the bulk solution is independent of $Re$ and we recover the amplitude equation obtained by \citet{Muraschko2015}
\begin{equation}
	\p_{Z}\varphi_{0}=0.
\end{equation}
However, for the boundary layer solution we find the scaling $m_ {bl}^{2}\sim{=}iRe\left(c-\overline{u}\right)$ at leading order in the large Reynolds limit. In this case, the amplitude equation for the boundary layer solution reduces to
\begin{equation}
	\p_{Z}\left(\varphi_{0}^{2}\left(c-\overline{u}\right)^{2}\right)=0.
\end{equation}
These results fail close to critical layers where $\left|c-\overline{u}\right|\ll1$.}

{Let us consider the  momentum flux  computed from the self-interaction of the upwardly propagating bulk solution of (\ref{eq:WKBDisp}), i.e. the one converging toward the inviscid solution when we take the limit $Re\to\infty$. If we assume $F\!r\left|c-\overline{u}\right|\ll1$ and $Re\left|c-\overline{u}\right|\sim \left(F\!r\left|c-\overline{u}\right|\right)^{-3}$ for every $z$, we recover the expression of equation (2.1) of \citet{Plumb1978} with $a_{1}=0$ :
\begin{equation}
	\overline{u^{\prime}w^{\prime}}\left(z\right)=\mathrm{sign}\left(c\right)\left|\varphi_{0}\left(z=0\right)\right|^{2} \exp\left\{-\frac{1}{F\!r^{3}Re}\int_{0}^{z}\frac{\mathrm{d}z^{\prime}}{\left(c-\overline{u}\left(z^{\prime}\right)\right)^{4}}\right\}.
\end{equation}
This expression fails close to critical layers where the scaling assumption $Re\left|c-\overline{u}\right|\sim\left(F\!r\left|c-\overline{u}\right|\right)^{-3}$ can not remain valid.}

\bibliographystyle{jfm}

\bibliography{jfm}

\end{document}